\documentclass[a4paper,twocolumn,11pt,accepted=2025-05-05]{quantumarticle}
\pdfoutput=1
\usepackage[utf8]{inputenc}
\usepackage[english]{babel}
\usepackage[T1]{fontenc}
\usepackage{amsmath}
\usepackage{hyperref}

\usepackage{tikz}
\usepackage{lipsum}

\usepackage{graphicx}
\usepackage{amsfonts}
\usepackage{amssymb}
\usepackage{amsmath}
\usepackage{dcolumn}
\usepackage[normalem]{ulem}
\usepackage{enumerate}
\usepackage{bbold}
\usepackage{cancel}
\usepackage{mathtools}
\usepackage{siunitx}
\usepackage{braket}
\usepackage{amsmath}
\usepackage{mdframed}
\usepackage{dsfont}
\usepackage[caption=false]{subfig}
\usepackage{soul}
\usepackage{orcidlink}

\definecolor{darkgreen}{RGB}{0, 159, 117}
\definecolor{navyblue}{rgb}{0.0, 0.0, 0.5}
\definecolor{darkolivegreen}{rgb}{0.33, 0.42, 0.18}
\definecolor{darktangerine}{rgb}{1.0, 0.66, 0.07}
\definecolor{darkspringgreen}{rgb}{0.09, 0.45, 0.27}

\usepackage{ mathrsfs }

\newcommand{\up}{\uparrow}
\newcommand{\down}{\downarrow}

\newcommand{\kket}[1]{\left.\ket{#1}\right>}
\newcommand{\bbra}[1]{\left<\bra{#1}\right.}

\begin{document}

\title{Stochastic resetting in discrete-time quantum dynamics: steady states and correlations in few-qubit systems}

\author{Sascha Wald}
\email{sascha.wald@coventry.ac.uk}
\affiliation{Centre for Fluid and Complex Systems, Coventry University, Coventry, CV1 2TT, United Kingdom}
\affiliation{$\mathbb{L}^4$ Collaboration \& Doctoral College for the Statistical Physics of Complex Systems, Leipzig-Lorraine-Lviv-Coventry, Europe}
\orcid{0000-0003-1013-2130}

\author{Louie Hong Yao}
\affiliation{Center for Soft Matter and Biological Physics, Virginia Tech, Blacksburg, VA 24061, USA}
\orcid{0000-0001-6910-2951}

\author{Thierry Platini}
\affiliation{Centre for Fluid and Complex Systems, Coventry University, Coventry, CV1 2TT, United Kingdom}
\affiliation{$\mathbb{L}^4$ Collaboration \& Doctoral College for the Statistical Physics of Complex Systems, Leipzig-Lorraine-Lviv-Coventry, Europe}

\author{Chris Hooley}
\affiliation{Centre for Fluid and Complex Systems, Coventry University, Coventry, CV1 2TT, United Kingdom}
\affiliation{$\mathbb{L}^4$ Collaboration \& Doctoral College for the Statistical Physics of Complex Systems, Leipzig-Lorraine-Lviv-Coventry, Europe}
\orcid{0000-0002-9976-2405}

\author{Federico Carollo}
\affiliation{Centre for Fluid and Complex Systems, Coventry University, Coventry, CV1 2TT, United Kingdom}
\affiliation{$\mathbb{L}^4$ Collaboration \& Doctoral College for the Statistical Physics of Complex Systems, Leipzig-Lorraine-Lviv-Coventry, Europe}
\maketitle

\begin{abstract}
 Time evolution in several classes of quantum devices is generated through the application of quantum gates.
 Resetting is a critical technological feature in these systems allowing for mid-circuit measurement and complete or partial qubit reset. 
 The possibility of realizing discrete-time reset dynamics on quantum computers makes it important to investigate the steady-state properties of such dynamics.
 Here, we explore the behavior of generic discrete-time unitary dynamics interspersed by random reset events.
 For Poissonian resets, we compute the stationary state of the process and demonstrate, by taking a weak-reset limit, the existence of ``resonances” in the quantum gates, allowing for the emergence of steady state density matrices which are not diagonal in the eigenbasis of the generator of the unitary gate. 
 Such resonances are a genuine discrete-time feature and impact on quantum and classical correlations even beyond the weak-reset limit.
 Furthermore, we consider non-Poissonian reset processes and explore conditions for the existence of a steady state. 
 We show that, when the reset probability vanishes sufficiently rapidly with time, the system does not approach a steady state. Our results highlight key differences between continuous-time and discrete-time stochastic resetting and may be useful to engineer states with controllable correlations on existing devices.
\end{abstract}

\section{Introduction}
A physical system subject to stochastic resetting is interrupted, at random times, by a reinitialization to a dedicated reset state \cite{evans2011diffusion}.
These kinds of processes have been extensively investigated in classical systems \cite{evans2011diffusion,Evans_2014,majumdar2015,pal_2015,PhysRevE.92.062148,mendez_2016,Fuchs_2016,Nagar_2016,Harris_2017,Maes_2017,Mag_2020,Monthus_2021,Biroli_2023} (see also, e.g., Refs.~\cite{evans2020stochastic,Gup22,Nag23} for comprehensive reviews) and show a variety of intriguing applications in stochastic algorithms, search as well as optimization problems~\cite{Tong07,PhysRevLett.113.220602,PhysRevE.92.062115,PhysRevLett.118.030603,valdeolivas2019random,Pal23,Blu22}. 
It has been shown that stochastic resetting may give rise to {\it universal probability laws}, such that the probability of certain observations in single realizations of the process are completely independent of the underlying dynamics, of the reset state and/or of the observed quantities themselves~\cite{Smi23,God23}.
Such universality is of fundamental importance since it allows for general predictions on a broad class of stochastic processes. 
Recently, there has also been increased interest in how stochastic resetting affects quantum dynamics~\cite{mukherjee2018quantum,rose2018spectral,Wald21,Perfetto_2021,Mag22,Per22,Kul23,Kul24,carollo_2024,Xhek22,Geh24}.
A wealth of different features have been observed, such as distinct node sampling statistics on complex networks~\cite{Wald21}, entanglement transitions in many-body systems~\cite{Xhek22},  generation of entangled steady states~\cite{Kul23,Kul24} or  of  correlations in noninteracting dynamics~\cite{Mag22}, and potential application to  quantum error correction~\cite{Geh24}.

\begin{figure}[t]
    \centering

    \includegraphics[width = \columnwidth]{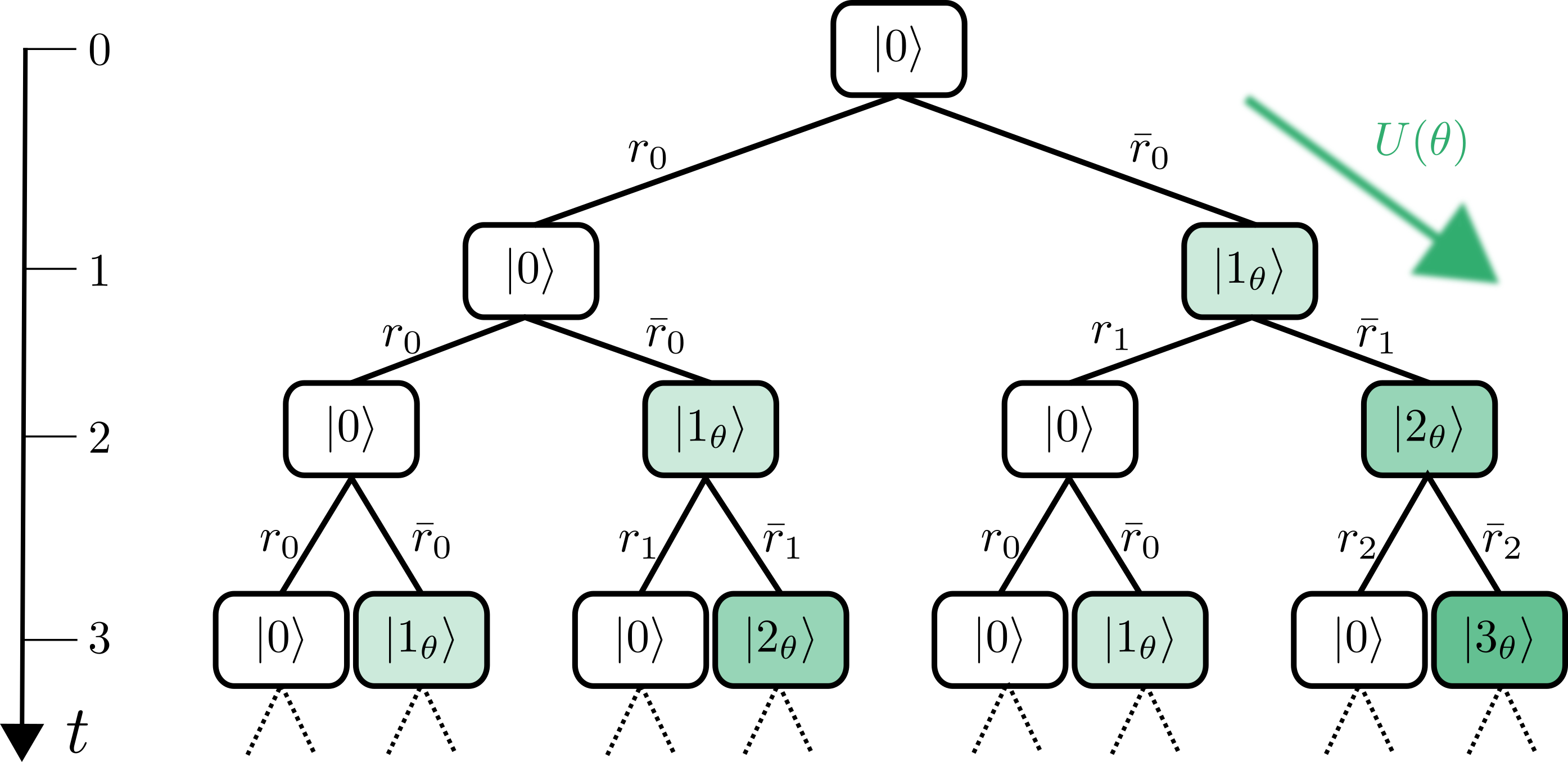}

    \caption{
        {\bf Ensemble of dynamical realizations of a discrete-time  evolution subject to stochastic  resetting.}
        A quantum system initialized in state $\ket{0}$ evolves according to a discrete-time unitary dynamics interspersed by reset events.
        The unitary dynamics explores the states $\ket{n_\theta}$, obtained by applying the unitary $U(\theta)$ $n$ times on the initial state. 
        Due to the presence of a reset process, at each discrete time, the system can either evolve unitarily or be reset to its initial state.
        The probability, $r_n$, with which a reset event occurs solely depends on the time elapsed since the previous reset. The unitary update instead takes place with probability $\bar{r}_n=1-r_n$. The presence of two possible dynamics at each time step allows one to visualize the ensemble of dynamical realizations via the binary tree sketched in the figure. }

    \label{fig:tree}

\end{figure}

Stochastic resetting in quantum systems has been mostly investigated in the presence of an underlying continuous-time dynamics~\cite{rose2018spectral,Perfetto_2021,Wald21,Haack_2021,Mag22,Sev23,Kul23,Per22,Kul24,PhysRevE.109.064150,carollo_2024}. 
However, discrete-time quantum dynamics is nowadays of particular interest because of its relevance to sequences of clocked gate operations on a qubit register, as used for the processing of quantum information in many modern quantum circuits \cite{Niel10}.
This interest is also rooted in the fact that any quantum operation (on a finite system) can be approximated via a  sequence of quantum gates (see Solovay-Kitaev theorem~\cite{Kit97,Daw05}). 
Many available quantum computers even allow for mid-circuit measurements and resetting, which permits reusing of  computational resources and can play an important role for scalable quantum error correction \cite{DeC23}. 
Despite such relevance, with the exception of some recent work on first-passage problems \cite{Das2022} and quantum hitting times \cite{Yin23} the interplay between reset processes and discrete-time quantum dynamics as well as the emergence of correlations due to collective resetting in these setups remains largely unexplored.

In this work, we investigate the emergence of steady states, and  the behavior of their quantum and classical correlations, in  discrete-time reset processes. 
In particular, we consider a unitary discrete-time dynamics, which is interspersed by random reset events (cf.~Fig.~\ref{fig:tree}) whose probability solely depends on the time elapsed since the last reset.
This setup covers a broad class of discrete-time reset processes including Poissonian (constant reset probability) and non-Poissonian ones.
For the former cases, we derive the steady state of the system, which always exists, and analytically compute its structure in the limit of vanishing reset probability.
Contrary to continuous-time dynamics, in which the steady state in such a  weak-reset limit is represented by an  ensemble which is diagonal in the Hamiltonian eigenbasis~\cite{Kul23}, we demonstrate that in our discrete-time setting off-diagonal contributions can survive the limit.
This occurs as a consequence of degeneracies, or as we call them here {\it resonances}, in the spectrum of the unitary quantum gate (analogous degeneracies were  also recently discussed in the context of 
stroboscopically measured continuous-time dynamics   \cite{Wang2024,Yin25b}). By means of two examples we show that these resonances strongly affect  quantum and classical steady-state correlations, even beyond the weak-reset limit. 
Finally, we  consider a  non-Poissonian stochastic resetting  and explore conditions for the existence of a  steady state. As we discuss, if the reset probability  decreases sufficiently rapidly with the time elapsed from the previous reset event, the system has a finite chance of not experiencing any reset  and a steady state may not exist. 

The rest of the paper is organized as follows.
In Sec.~\ref{sec:theory} we introduce the discrete-time process we consider in this work and provide analytical considerations on the evolution of the quantum state of the system.
We then focus on the Poissonian reset and discuss key differences between discrete-time and continuous-time reset processes. 
In Sec.~\ref{sec:2qb-poisson}, we consider a two-qubit system evolving under a Poissonian protocol. 
We consider two cases: (i) a noninteracting quantum gate and (ii) an interacting dynamics inspired by a paradigmatic entangling circuit.
In both cases we numerically explore steady-state correlations and observe the emergence of resonances.
Sec.~\ref{sec:nm} is concerned with the existence of the steady state in the case of non-Poissonian resets.
In Sec.~\ref{sec:conclusion}, we provide a discussion and conclusions. Details on certain calculations are outlined in the appendices.

\section{Discrete-Time Resetting Dynamics}
\label{sec:theory}

\subsection{The model}
We consider a generic quantum system that evolves in discrete time via the application of a quantum gate $U(\theta) = \exp(i \theta \mathscr{H})$.
Here, $\mathscr{H}$ is a Hamiltonian-like generator for the unitary gate and $\theta$ is a tunable real parameter.  We consider both $\mathscr{H}$ and $\theta$ to be dimensionless.
Given an initial quantum state $\ket{0}$, the unitary dynamics explores the states $\ket{n_\theta} = U^n(\theta) \ket{0}$, which are in general not mutually orthogonal, i.e., $\braket{n_\theta|m_\theta} \neq \delta_{nm}$.
In addition to this deterministic evolution, the quantum system is also subject to a reset process which, stochastically in time, reinitializes the system state to  $\ket{0}$ (cf.~Fig.~\ref{fig:tree}).
More precisely, the probability of a reset event occurring, at any given discrete time, is denoted as $r_n\in[0,1]$ and is assumed to depend solely on the number  $n$ of unitary update events since the last reset took place. 
An illustrative representation of all possible realizations associated with this stochastic resetting dynamics is shown in Fig.~\ref{fig:tree}. Each branch of the binary tree sketched in the figure corresponds to a single realization, or quantum trajectory, of the dynamics in, e.g., a numerical simulation or an experimental run.
The above setup encompasses ``Poissonian" resetting, characterized by $r_n = r $, as well as deterministic resetting protocols, in which $r_n = \delta_{n,\ell}$ for a given $\ell$ \cite{Yin23}.

The quantum state describing the evolution of the system on average can be obtained by taking into account all possible dynamical realizations, weighted with their corresponding probabilities.
At any discrete time $t$, its density matrix assumes the form 
\begin{align}\label{eq:varrho}
 \varrho(t) = \sum_{n=0}^t P_n(t) \ket{n_\theta}\!\bra{n_\theta},
\end{align}
with $P_n(t)$ being the probability to find the system in state $\ket{n_\theta}$ at time $t$. 
For $n\geq 1$, the probability $P_n(t)$  satisfies the recursion relation
\begin{align}
 P_n(t) = (1-r_{n-1}) P_{n-1}(t-1)\, ,
 \label{eq:recPn}
\end{align}
which follows from the fact that the system can reach the state $\ket{n_\theta}$ at time $t$, only if it is in state $\ket{{(n-1)}_\theta}$ at time $t-1$. 
On the other hand, since the state $\ket{0}$ can be reached, at time $t$, from any state $\ket{n_\theta}$ with $0\leq n < t$, due to a reset event, we can write  
\begin{align}
 P_0(t) = \sum_{n=0}^{t-1} r_n P_n(t-1).
 \label{eq:P0}
\end{align}
The relations in~(\ref{eq:recPn}) and~(\ref{eq:P0}) can be cast into a simple matrix-vector equation for the probability.
We indeed have $P(t+1)=R(t)P(t),$ with ${ P}(t)=[P_0(t),P_1(t),\hdots , P_t(t)]^T$ and ${ R}(t)$ being the  sparse rectangular matrix of size $(t+2)\times (t+1)$, 
\begin{align}
    R(t) =
    \begin{pmatrix}
    r_0 & r_1 & \hdots & r_{t}\\
    1-{r}_0& 0 & \hdots & 0\\
    0 & 1-{r}_1 & \hdots & 0\\
    \vdots &\ddots & \ddots&\vdots\\
    0 & \hdots & 0 & 1-{r}_t
    \end{pmatrix}.
\end{align}
The relation \eqref{eq:recPn} further allows us to express all $P_n(t)$ (with $t\geq n$) through the distribution $P_0(\tau)$ as 
\begin{align}
 P_n(t) = P_0(t-n)\prod_{j=0}^{n-1} (1-r_j).
 \label{eq:Prec}
\end{align}
Hence, full knowledge of the time-dependent probability $P_0(t)$, which equals the probability of a reset event occurring at time $t$, completely determines the problem. 

Note that~\eqref{eq:varrho}, together with the relations~\eqref{eq:P0} and~\eqref{eq:Prec}, is equivalent to a {\it last renewal equation} (see, e.g., Refs.~\cite{evans2020stochastic,Mag22,Das2022}), typically exploited in the framework of reset processes. 
Indeed, since the  system can only be in state $\ket{n_\theta}$ at time $t$ if a reset to state $\ket{0}$ occurred at time $t-n$ without any subsequent resetting occurring, one can write 
\begin{align}
\begin{split}
 \varrho(t) &= P_t(t) \ket{t_\theta}\!\bra{t_\theta}\\
  &\quad+ \sum_{m=1}^tP_{t-m}(t)\ket{(t-m)_\theta}\!\bra{(t-m)_\theta} .
\end{split}
\label{eq:renew}
\end{align}
In the above equation, $P_t(t)$ is the probability that the system evolves up to time $t$ without being subject to any reset event.
As apparent from \eqref{eq:Prec}, $P_{t-m}(t)$ denotes instead the probability that a reset event occurred at time $t-m$ and the system evolved thereafter without experiencing  any further reset. 

As we shall discuss in detail below, the setup introduced in this section does not guarantee the existence of a steady state, $\varrho_{\rm ss}=\lim_{t\to\infty} \varrho(t)$. Cases that do not yield a steady state include fully unitary dynamics ($r_n =0$) and deterministic resetting ($r_n =\delta_{n,\ell}$).  There are, however, also more subtle instances in which the reset probability $r_n$, while never zero, vanishes too rapidly with $n$ for a steady state to be reached.  In these cases, the dynamics remains periodic at arbitrarily long times.

\subsection{The Poissonian case}
The relations obtained in the previous section provide a way to investigate the discrete-time resetting dynamics, at least numerically. To make analytical progress and to shed light on certain features which are specific to discrete-time reset processes, we focus, for the moment,  on the case of Poissonian resets. 
In this setting, the reset probability $r_n$ is constant, $r_n = r >0$, which guarantees that the process does not depend on the history of the reset events. The evolution of the quantum state can thus be written in terms of a Markovian discrete-time quantum master equation as 
\begin{align}
 \varrho(t+1) = r \ket{0}\!\bra{0} + (1-r) U(\theta)\varrho(t)U(\theta)^\dagger \ .
\end{align}
No such master equation is attainable for the generic process discussed in the previous section. 
For Poissonian resetting, the underlying stochastic process is a Bernoulli chain, such that $P_0(\tau > 0) = r$, which further allows us to evaluate
\begin{align}
  P_n(t) = (1-r)^{n} (r+(1-r)\delta_{n,t}).
\end{align}
In this regime, \eqref{eq:renew} simplifies to 
\begin{align}
 \varrho(t) &= (1-r)^t\ket{t_\theta}\!\bra{t_\theta} +
 r\sum_{n=0}^{t-1} (1-r)^{n} \ket{n_\theta}\!\bra{n_\theta}\, .
\end{align}
In the above equation, the first term describes the purely unitary evolution, weighted with its probability, and each term in the sum accounts for a reset event at any previous time followed by a suitable number of applications of the unitary gate without further resets. In the case of Poissonian resetting, the system always approaches a steady state, for long times, given by 
\begin{align}
\label{eq:steady-pois}
\varrho_{\rm ss} = r\sum_{n=0}^{\infty} (1-r)^{n} \ket{n_\theta}\!\bra{n_\theta}.
\end{align}

\subsection{The limit of weak Poissonian resetting}
With the expression of the steady state (cf.~(\ref{eq:steady-pois})) at hand, we can now explore the behavior of the system in the weak-resetting limit. That is, we consider the steady state $\varrho_{\rm ss}$ in \eqref{eq:steady-pois} and take the limit $r\to0$. 
This limit allows us to analytically show a peculiar feature, which can be only observed in discrete-time resetting dynamics, and shed light on how resetting mixes different branches of the tree in Fig.~\ref{fig:tree} into the average quantum state of the system.

The weak-resetting limit can be conveniently analyzed by considering the eigenstates $\ket{e_i}$ of the generator, $\mathscr{H} \ket{e_i} = \lambda_i \ket{e_i}$.
Note that:\ (i) although the eigenstates $\ket{e_i}$ do not depend on the parameter $\theta$, they are eigenvectors of the quantum gate $U(\theta)$ too, since $[U(\theta),\mathscr{H}] = 0$; (ii) the eigenvalues $u_i(\theta)$ of the quantum gate do depend on $\theta$; (iii) the eigenvalues $u_i(\theta)$ satisfy $u_i(\theta)^* = 1/u_i(\theta)$, since $U(\theta)$ is a unitary operator.
Exploiting these eigenstates and 
performing the summation over $n$ in (\ref{eq:steady-pois}), we find
\begin{align}
\varrho_{\rm ss}&= \sum_{ij}\frac{r \braket{e_i|0 }\!\braket{0|e_j}}{1-(1-r) u_j(\theta)/u_i(\theta)}  \ket{e_i}\!\bra{e_j}.
\label{eq_weak_reset_limit}
\end{align}

In the limit $r\to0$, we therefore observe that if $u_j(\theta) \neq u_i(\theta)$ the corresponding component of the quantum state vanishes, thus giving rise to a diagonal state in the generator eigenbasis $\ket{e_i}$. This convergence to a {\it diagonal ensemble} in the weak-reset limit is exactly what one would expect from known results on stochastic resetting in continuous time  \cite{Kul23}.

In our discrete-time setup, however, different behavior can emerge. If $u_j(\theta) = u_i(\theta)$, off-diagonal components in the generator eigenbasis $\ket{e_i}$ can survive the weak-reset limit. Considering that we have 
$u_j(\theta) = e^{i\theta \lambda_j}$, with $\lambda_j$ being the eigenvalues of $\mathscr{H}$, the relation $u_j(\theta) = u_i(\theta)$ can be met whenever the {\it resonance condition\/} 
$\theta(\lambda_i - \lambda_j) = 2\pi k$, with $k$ an integer, is satisfied. As a consequence of this resonance, the quantum state in the weak-reset limit can display nonvanishing coherence in the eigenbasis of $\mathscr{H}$.
For continuous-time dynamics, such off-diagonal contributions can only occur when the system Hamiltonian features degenerate eigenspaces. 
Even in such cases, the degeneracy in $\mathscr{H}$ allows to find a suitable eigenbasis of the generator in which off-diagonal terms vanish.
On the other hand, for discrete-time dynamics the off-diagonal contributions are more general and cannot be removed by a change of basis, since the degeneracies of $\mathscr{H}$ and $U(\theta)$ are decoupled when the resonance condition is met.
As discussed below, these resonances are not only relevant in the $r\to 0$ limit but leave important signatures also at finite values of $r$. 

Summarizing the above considerations, 
 we find that the steady state is characterized by 
%
%
\begin{align}
\lim_{r\to 0} 
\bra{e_i}\varrho_{\rm ss}\ket{e_j} =
 \begin{cases}
    \left| \braket{e_i|0 }\right|^2, \ { \rm if}\ i = j\\[.25cm]
    \braket{e_i|0 }\!\braket{0|e_j}, \ {\rm if}\ \theta\frac{\lambda_i - \lambda_j}{2\pi} = k \\[.25cm]
    0, \ {\rm else}
 \end{cases} 
 \label{eq:rho_r0}
\end{align}
%
%
with $k$ being a non-zero integer. Note that, having $\theta(\lambda_i - \lambda_j) = 0$ for all $\theta$,  indicates degenerate eigenvalues for $\mathscr{H}$. The matrix elements in \eqref{eq:rho_r0} solely depend on the decomposition of the initial state into the eigenbasis $\ket{e_i}$. 
The weak-reset limit serves as a tool to suppress non-resonant contributions to the steady state.
The resonant contributions survive this limit since they are ``phase protected" from the dynamics, as $U(\theta) \ket{e_i}\bra{e_j} U(\theta)^\dagger = \ket{e_i} \bra{e_j}$ if the resonance condition between $\lambda_i$ and $\lambda_j$ is met.

\section{Two-Qubit Systems with Poissonian resetting }
\label{sec:2qb-poisson}
We now consider two concrete example systems each consisting of two qubits. This allows us to explore in detail the impact of the resonances discussed above on properties of the quantum state. 
 In particular, 
we are interested in investigating quantum and classical correlations in the steady state generated by the reset process. 

To this end, we focus on different measures of correlations. The first is the correlation between the $z$-components of the spins, defined as 
\begin{align}
 C \equiv \langle \sigma_z^{(1)} \sigma_z^{(2)}\rangle - \langle \sigma_z^{(1)}\rangle \langle \sigma_z^{(2)}\rangle \ .
\end{align}
Here, the superscript indicates the qubit to which the observable belongs, i.e., $\sigma_z^{(1)} = \sigma_z \otimes \mathds{1}$ and $\sigma_z^{(2)} = \mathds{1} \otimes \sigma_z$, $\sigma_z$ is the $z$ Pauli matrix and $\langle \mathcal{O} \rangle = \operatorname{tr}( \mathcal{O} \varrho_{\rm ss})$ is the expectation value of the operator $ \mathcal{O}$ in the steady state. 

Another quantity we consider is the steady-state entanglement. For a two-qubit system, the latter is  fully characterized by the concurrence $\mathcal{C}$, which is an entanglement monotone for mixed states \cite{Hill97}.
It is computed  as
\begin{align}
 \mathcal{C}(\varrho )\equiv \max(0,\mu _{0}-\mu _{1}-\mu _{2}-\mu_{3})\, ,
\end{align}
where $\mu_i$, for $i = 0,1,2,3$, are the eigenvalues in descending order of the matrix $( \varrho^{1/2} \tilde{\varrho} \varrho^{1/2} )^{1/2}$, with $\tilde{\varrho} = (\sigma_y\otimes \sigma_y) \varrho^* (\sigma_y\otimes \sigma_y)$~\cite{Hill97}, and $\sigma_y$ is the $y$ Pauli matrix.

Finally, we quantify quantum correlations beyond entanglement \cite{Bera18}, exploiting the so-called {\it local quantum uncertainty} (LQU),  closely related to quantum discord~\cite{Giro13}. We consider such a quantity 
since it can be straightforwardly computed; in particular, unlike in the case of quantum discord, no optimization procedure is required.
The LQU is defined as
\begin{align}
 \mathcal{U} = 1 - \nu_{\rm max},
\end{align}
with $\nu_{\rm max}$ being the largest eigenvalue of the $3\times3$ symmetric matrix $W$, whose elements are 
$W_{ij} = \operatorname{tr}(\sqrt{\varrho} \sigma_i^{(1)}  \sqrt{\varrho} \sigma_j^{(1)})$, with $\sigma_i$ for $i,j=x,y,z$ referring to the Pauli matrices.

In the following, we study the impact of a Poissonian stochastic resetting on two different types of unitary dynamics, a noninteracting and an interacting one.

\subsection{Noninteracting two-qubit system}
\label{ssec:2qb-non}
\begin{figure*}[t]
 \centering
 \includegraphics[width = .5\columnwidth]{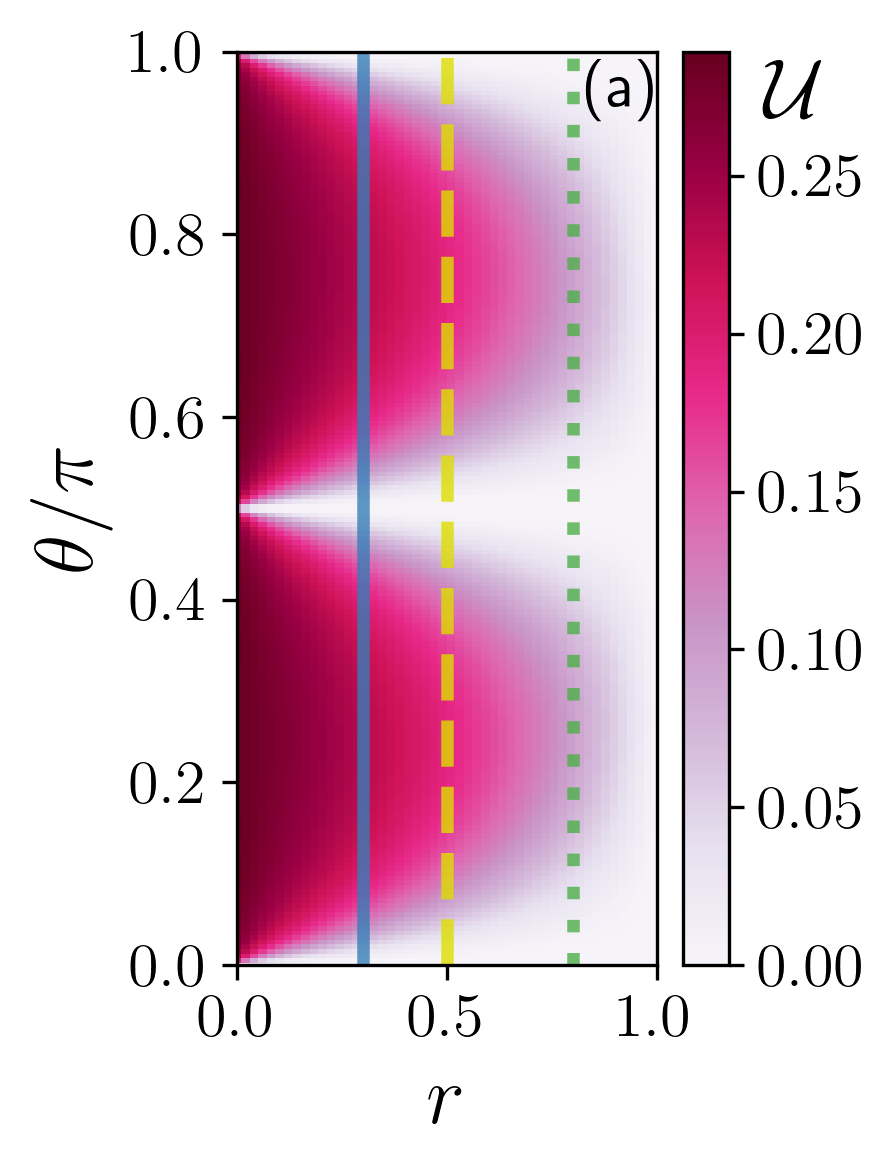}
 \includegraphics[width = .49\columnwidth]{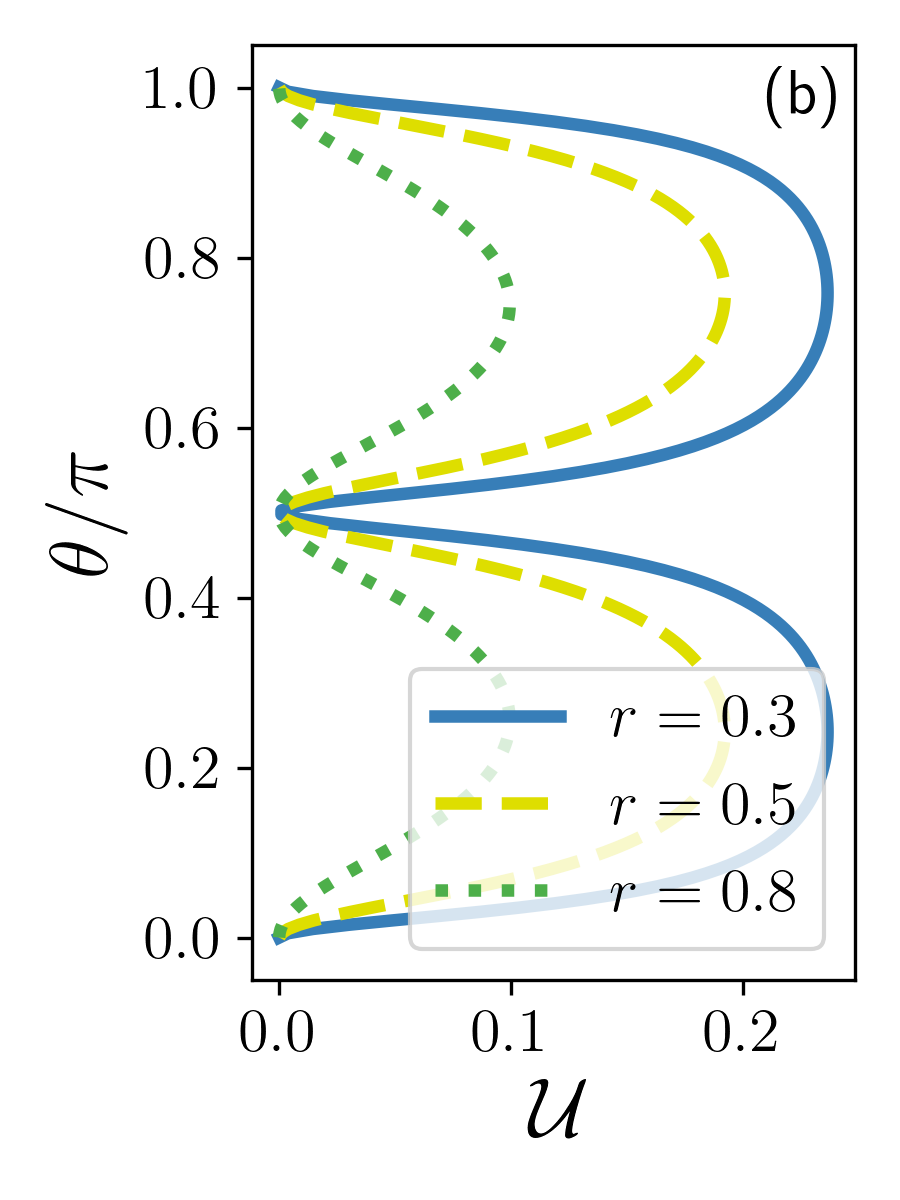}
 \includegraphics[width = .5\columnwidth]{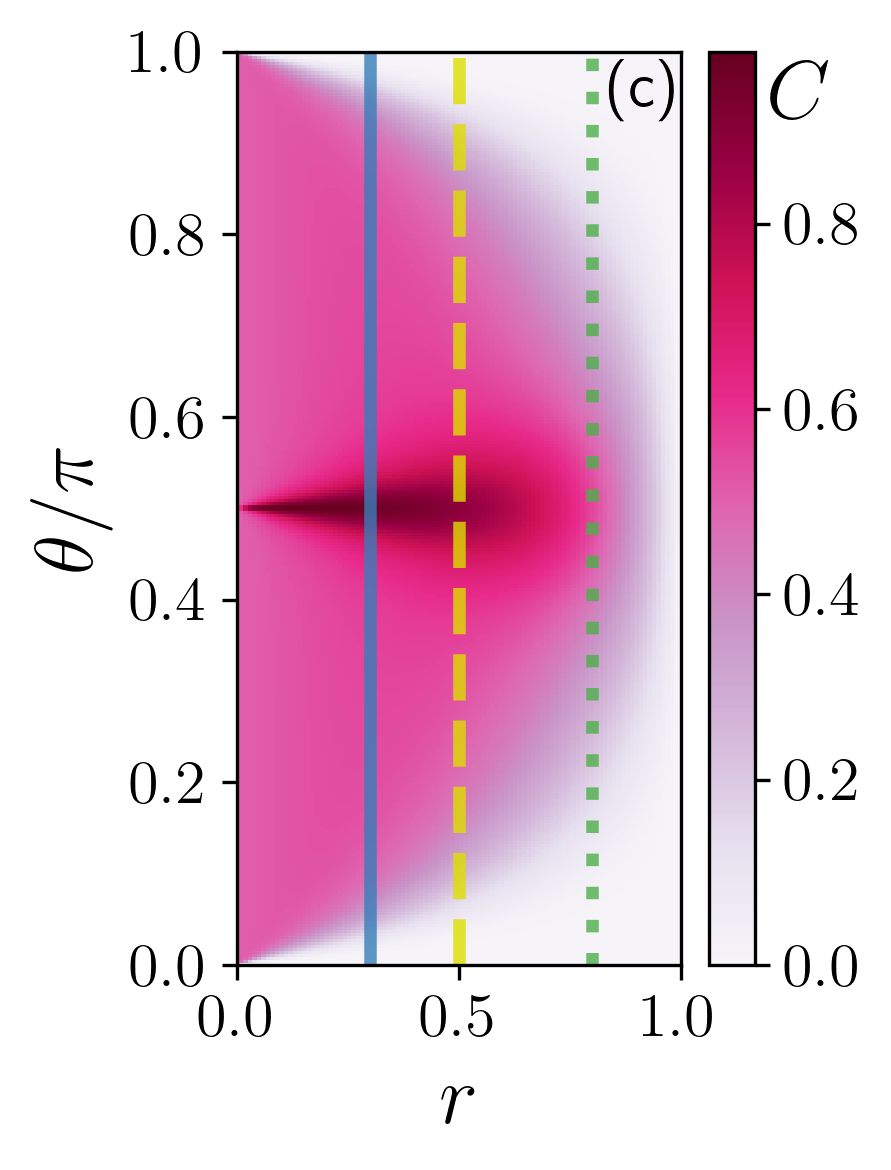}
 \includegraphics[width = .49\columnwidth]{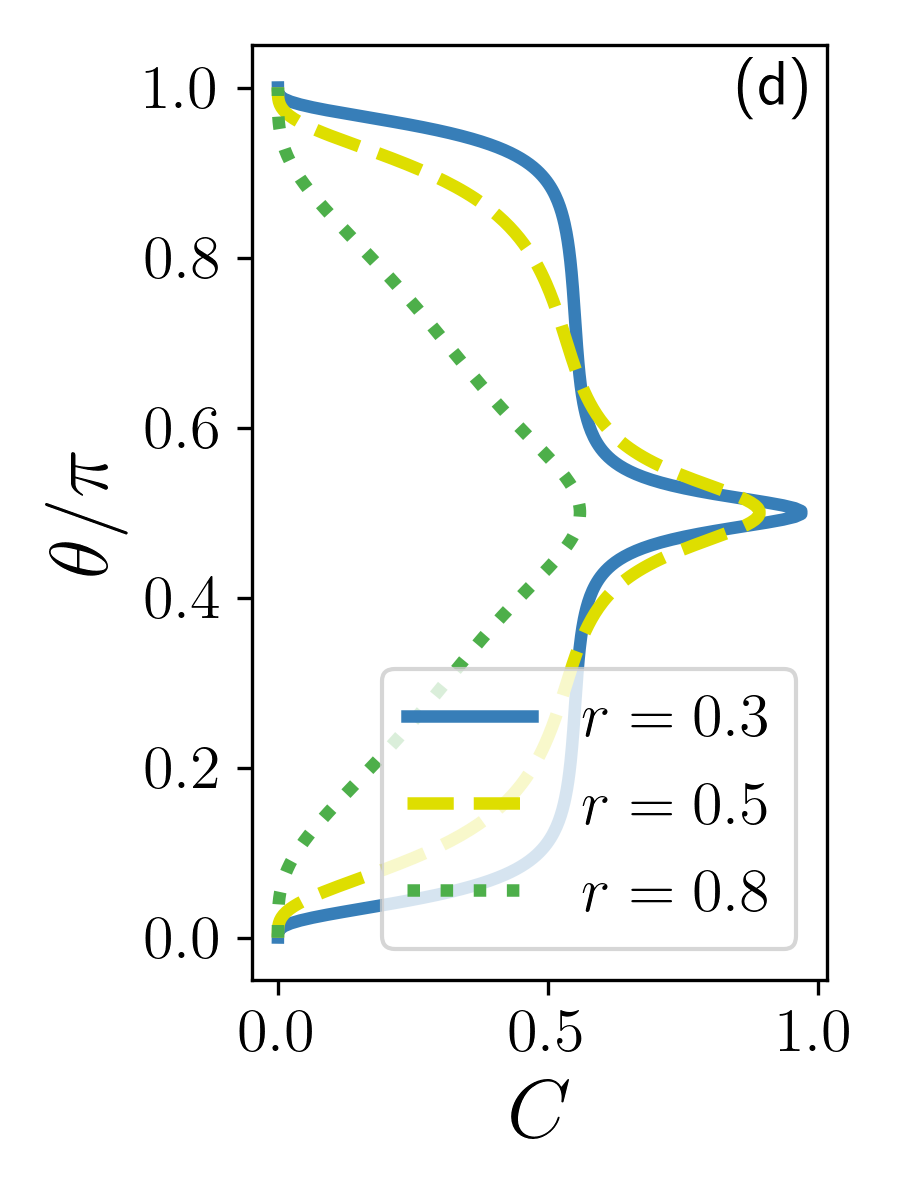}
 \caption{{\bf Steady-state correlations for noninteracting qubits.}  (a) Steady-state LQU as a function of $\theta$ and of the reset probability $r$. Nonzero values of the LQU show that the stochastic resetting generally induces quantum correlations. (b) Behavior of the LQU for specific values of the reset probability, $r=0.3,0.5,0.8$, highlighted also in panel (a). Panels (c-d) show analogous results as in panel (a-b) but concerning  the correlation $C$.  }
 \label{fig:non}
\end{figure*}
We first consider a system of two qubits undergoing a noninteracting unitary dynamics. The latter is generated by a Hamiltonian-like operator of the form 
\begin{align}
 \mathscr{H}_{\rm NI} = \sigma_x^{(1)} + \sigma_x^{(2)}
 =  \left(
\begin{array}{cccc}
 0 & 1 & 1 & 0 \\
 1 & 0 & 0 & 1 \\
 1 & 0 & 0 & 1 \\
 0 & 1 & 1 & 0 \\
\end{array}
\right), 
\label{eq:NI}
\end{align}
which can be seen as a coherent driving applied to the qubits.   
Here, we have used the $z$ basis $\{ \ket{\down},\ket{\up}\}$ to represent the operator.
As discussed above and shown by \eqref{eq:rho_r0}, steady-state properties  strongly depend on the spectrum of the operator $\mathscr{H}_{\rm NI}$ and on the value of $\theta$.
The eigenvalues are $\lambda_0 = -2$, $\lambda_1 = 0$ (with degeneracy $2$) and $\lambda_2 = 2$, with the corresponding eigenvectors
\begin{align}
\ket{e_{0}} &=
\ket{-}\otimes\ket{-}
\\[.2cm]
\ket{e_1^{1}} &=\ket{-}\otimes\ket{+}
\\[.2cm]
\ket{e_1^{2}} &=
\ket{+}\otimes\ket{-}
\\[.2cm]
\ket{e_2} &=
\ket{+}\otimes\ket{+}.
\end{align}
Here, $\ket{\pm} = (\ket{\up} \pm \ket{\down})/\sqrt{2}$ are the eigenvectors of $\sigma_x$.
If the parameter $\theta\neq \pi/2,\pi$, then there are no resonances in the sense introduced in the previous section (see discussion after \eqref{eq_weak_reset_limit}).
By choosing as initial state $\ket{0} = \ket{\up \up}$ and using \eqref{eq:rho_r0}, we find, in the weak-reset limit,
\begin{align}
 \lim_{r\to0}\varrho_{\rm ss} =\frac{1}{8}
 \begin{pmatrix}
    3&0&0&-1\\
    0&1&1&0\\
    0&1&1&0\\
    -1&0&0&3
 \end{pmatrix}\, .
\end{align}
This state features correlations in the $z$ direction, as witnessed by a finite value $C = 0.5$. It also features quantum correlations as captured by the LQU ($\mathcal{U} \approx 0.29$) even though the state is not entangled, i.e., $\mathcal{C} = 0$. 
These quantum correlations are here strictly related to the existence of a degenerate eigenvalue of $\mathscr{H}_{\rm NI}$.
In the absence of degeneracy (and of resonances), the state would be
a statistical mixture of the eigenstates of a noninteracting Hamiltonian-like generator $\mathscr{H}_{\rm NI}$, and would thus be equivalent to a classical state.

We now explore the impact of the previously mentioned resonances on the considered steady-state properties. 
First, for $\theta = \pi$ all eigenstates of $U$  are simultaneously resonant, i.e., $e^{i \pi \lambda_j} = 1$. This case is trivial since the parameterized gate $U(\pi) =\mathds{1}$ reduces to the identity. 
For $\theta = \pi/2$, on the other hand, only the eigenvalues $\lambda_0 = 2$ and $\lambda_2 = -2$ are in resonance as $e^{i\pi\lambda_0/2} = e^{i\pi\lambda_2/2}$.
In this case, it is straightforward to calculate 
\begin{align}
 \lim_{r\to0}\varrho_{\rm ss} =
 \begin{pmatrix}
  \frac{1}{2}&0&0&0\\
  0&0&0&0\\
  0&0&0&0\\
  0&0&0&\frac{1}{2}\\
 \end{pmatrix}\, ,
\end{align}
which can also be understood from the fact that the for $\theta = \pi/2$ the dynamics simply consists of a classical spin flip dynamics. 

In Fig.~\ref{fig:non}, we show the LQU and the spin-spin correlation in the steady state,  as a function of $\theta$ and of the reset probability $r$ (see Appendix~\ref{app:numerics} for details on the calculation of these quantities).
The spin-spin correlation also admits an analytical expression, 
\begin{align}
C(r,\omega)=
\frac{1+f(r,4\omega)}{2}
-
\left[f(r,2\omega)\right]^2\, ,
\end{align}
where the function $f$ is specified in Appendix~\ref{app:correlator}.
We see from Fig.~\ref{fig:non} that the steady state shows quantum and classical correlations in large portions of the parameter space.
Correlations are largest along the classical resonance where the LQU vanishes.
From this example, we see that while the condition for the existence of the resonances is derived in the weak-reset probability limit, this resonance still significantly affects properties of the steady state also for finite reset probability.

\subsection{Entangling unitary dynamics}
\label{ssec:Tgate}
\begin{figure}[h!]
\includegraphics[width= .9\columnwidth]{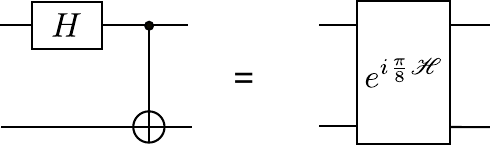}
\caption{{\bf Entangling circuit.}
Quantum circuit that generates a maximally entangled Bell state from the two qubit product state $\ket{\up\up}$.
We derive the generator $\mathscr{H}$ of this circuit and use it to define a parameterized quantum gate.}
\label{fig:2qbits}
\end{figure}
\begin{figure*}[t]
 \centering
 \includegraphics[width=.49\columnwidth]{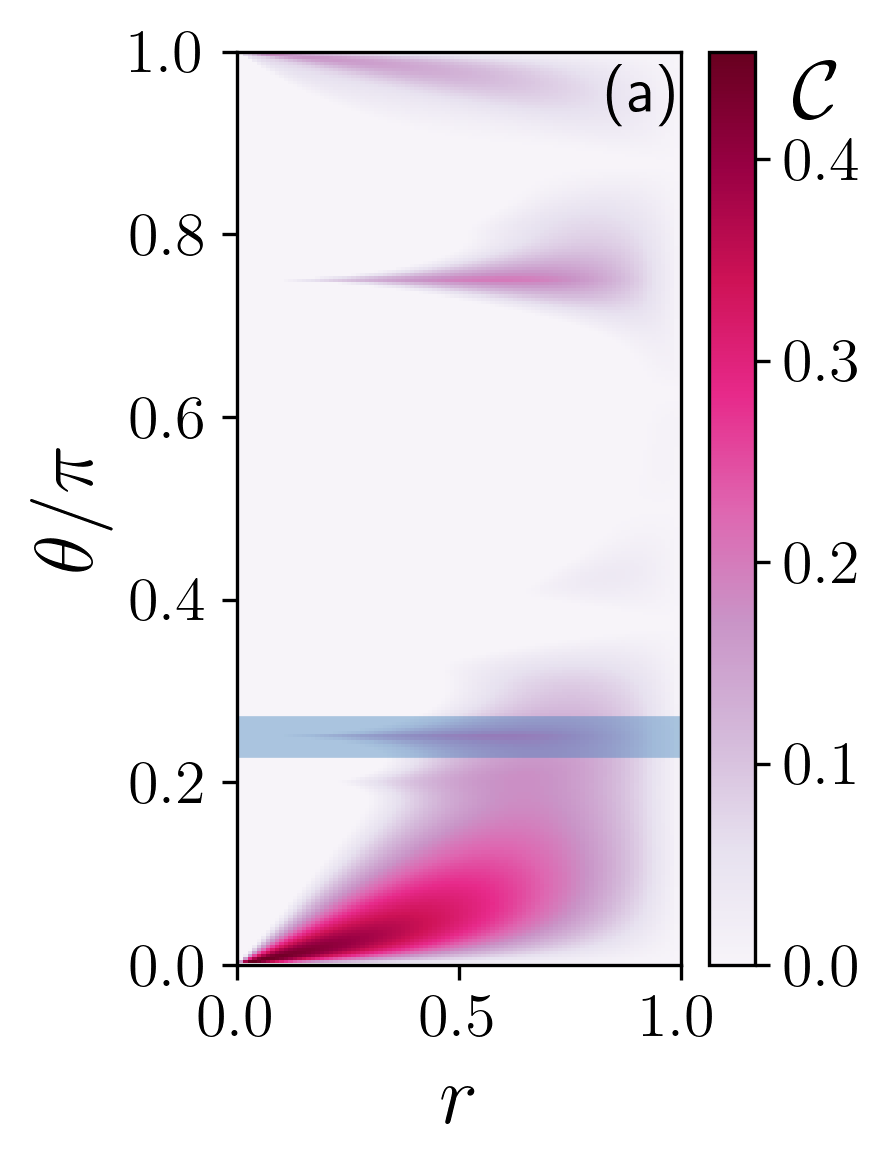}
 \includegraphics[width=.49\columnwidth]{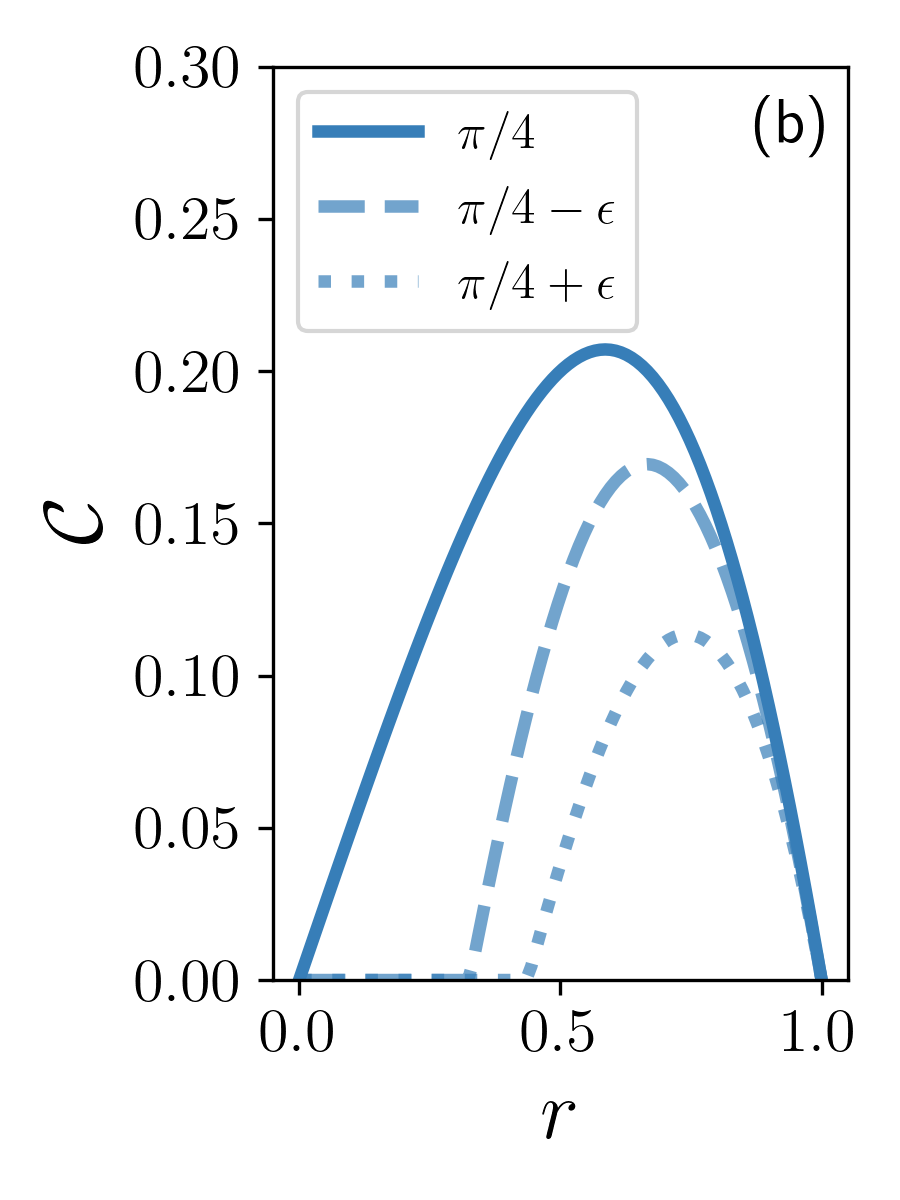}
 \includegraphics[width=.49\columnwidth]{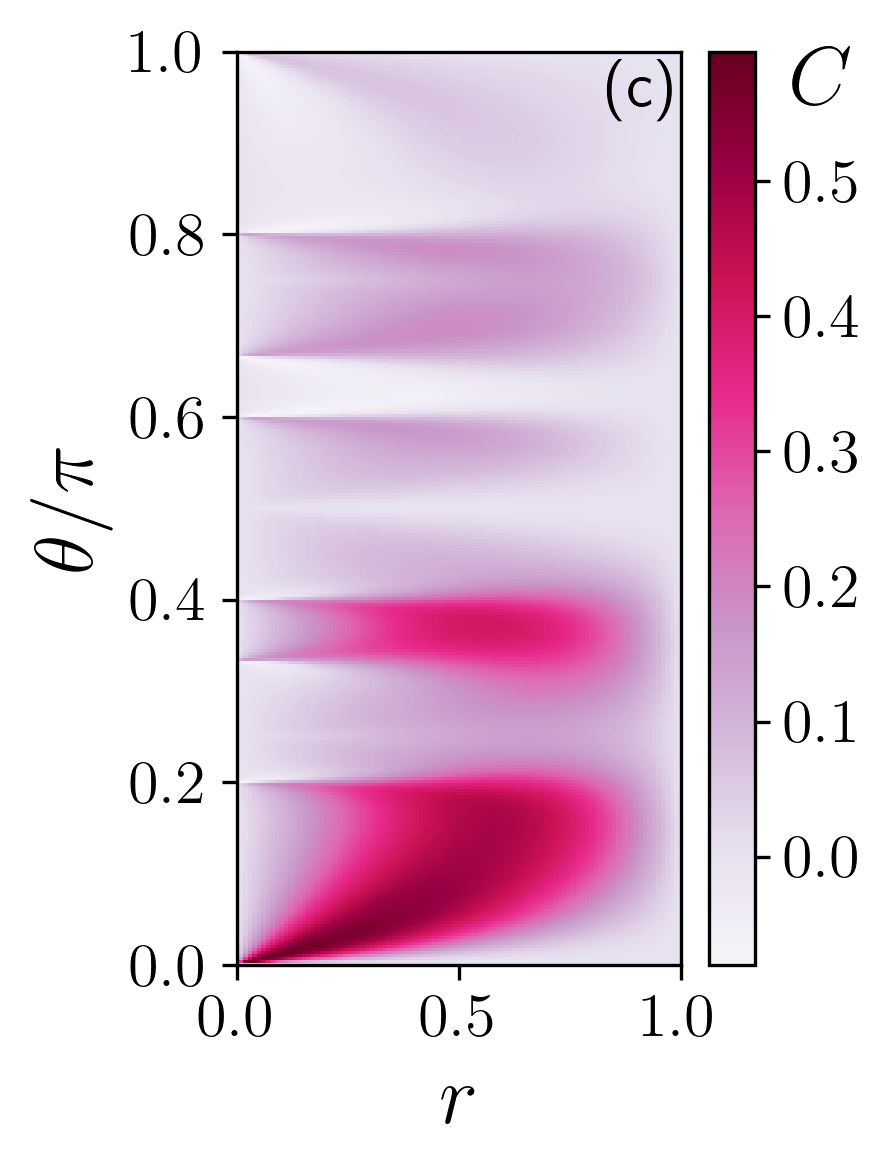}
 \includegraphics[width=.49\columnwidth]{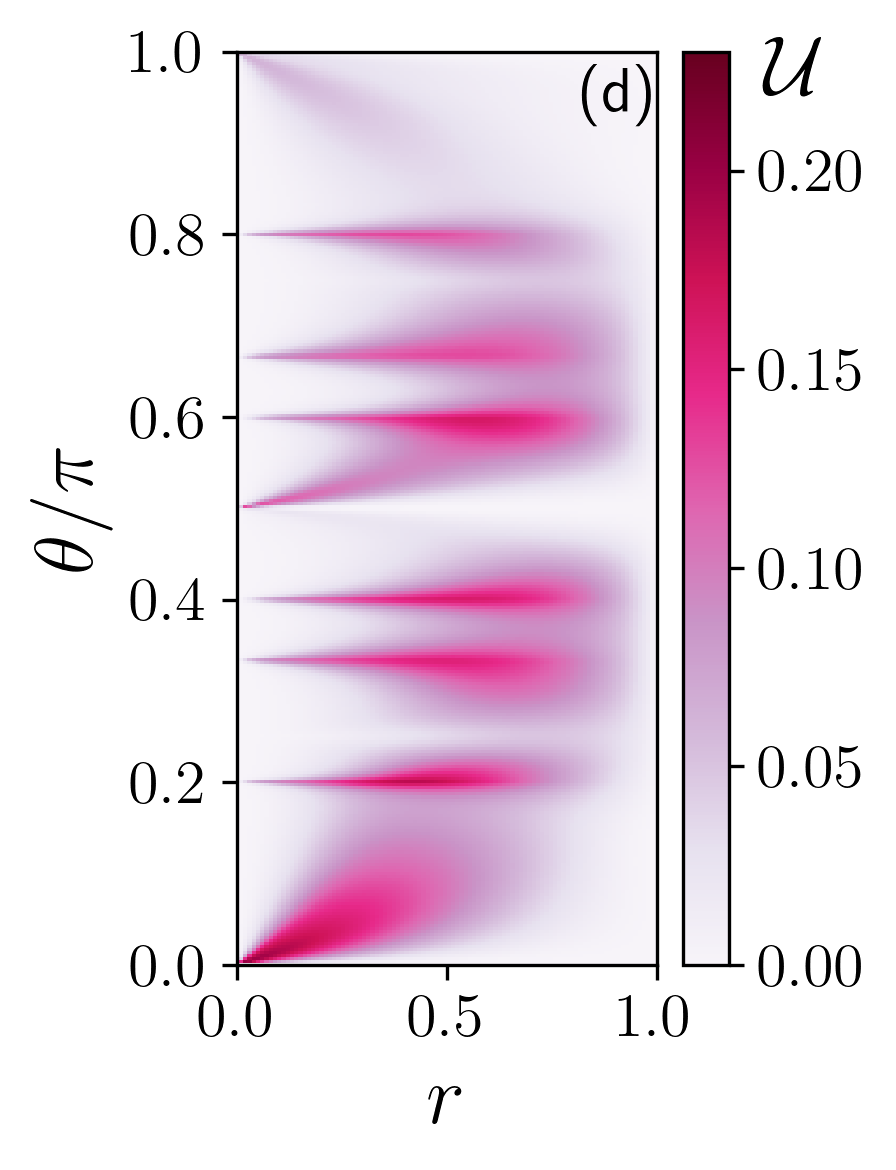}

 \caption{ {\bf Steady-state correlations for the entangling dynamics.}
 (a) Entanglement between the two qubits, as measured by the concurrence, as a function of $\theta$ and of the reset probability $r$. (b) Plot of the concurrence, as a function of the reset probability, for the values $\theta=\pi/4,\pi/4\pm\epsilon$, with $\epsilon=0.01$ (see shaded region in panel (a)).  (c) Classical correlations in the $z$ magnetization for the two qubits, as a function of $\theta$ and the reset probability. (d) LQU for the two-qubit system, as a function of $\theta$ and the reset probability. The existence of resonances is clearly highlighted by the presence of sharp horizontal features in panels (a),(c),(d).
 }
 \label{fig:Tgate}
\end{figure*}
In this section, we consider a unitary gate $U(\theta)$ which can generate entanglement between the two qubits. In this way, we can explore how stochastic resetting competes with a discrete-time entangling evolution. 
The unitary dynamics we consider takes inspiration from the well-known quantum circuit featuring a Hadamard gate ($H$) and a CNOT gate (see Fig.~\ref{fig:2qbits}). This circuit is used to generate a maximally entangled Bell state starting from the product state $\ket{\up\up}$,
\begin{align}
 \ket{\up\up}\quad  \stackrel{H}{\longrightarrow}\quad \frac{\ket{\up\up} + \ket{\down \up}}{\sqrt{2}} \quad
 \stackrel{{\rm CNOT}}{\longrightarrow} \quad \frac{\ket{\up\up} + \ket{\down \down}}{\sqrt{2}}.
\end{align}
To define a unitary dynamics that gradually builds up entanglement, we introduce a parameterized quantum gate $U(\theta)$ that is built from the generator $\mathscr{H}$ of the quantum circuit defined via the relation ${\rm CNOT} (H \otimes \mathds{1}) = \exp(i \pi \mathscr{H} /8 )$.
The Hamiltonian-like generator thus reads
\begin{align}
 \mathscr{H}\hspace{-.1cm} =\hspace{-.1cm}
 \begin{pmatrix}
 2-\sqrt{2} & 2-\sqrt{2} & -i-\sqrt{2} & i-\sqrt{2} \\
 2-\sqrt{2} & 2-\sqrt{2} & i-\sqrt{2} & -i-\sqrt{2} \\
 i-\sqrt{2} & -i-\sqrt{2} & \sqrt{2}+2 & \sqrt{2}+2 \\
 -i-\sqrt{2} & i-\sqrt{2} & \sqrt{2}+2 & \sqrt{2}+2 
 \end{pmatrix}
\end{align}
and we use it to construct the unitary  gate $U(\theta) = \exp(i \theta \mathscr{H})$.

Before exploring  steady-state properties, we discuss the spectral decomposition of the operator $\mathscr{H}$. 
Its eigenvalues are nondegenerate and are given by
$\lambda_0 = -2$, $\lambda_1 = 0$, $\lambda_2 =2$ and $\lambda_3 = 8$. 
The corresponding eigenvectors read
\begin{align}
 \ket{e_{0}}  &=
 \frac{i\ket{\up} + \ket{\down}}{\sqrt{2}} \otimes \ket{-}\\[.2cm]
\ket{e_1} &=
\frac{(\sqrt{2}+1) \ket{\up} + \ket{\down}}{\sqrt{2}\sqrt{2+\sqrt{2}}} \otimes  \ket{+}\\[.2cm]
\ket{e_2} &=
\frac{-i\ket{\up} + \ket{\down}}{\sqrt{2}} \otimes \ket{-}\\[.2cm]
\ket{e_3} &=
\frac{(1-\sqrt{2})\ket{\up} + \ket{\down}}{\sqrt{2} \sqrt{2-\sqrt{2}}}\otimes \ket{+} \ .
\end{align}
Considering values of $\theta$ which avoid resonances and taking the limit $r\to 0$, we find the steady state
\begin{align}
\lim_{r\to0}\varrho_{\rm ss}
 &=
\frac{1}{16}\left(
\begin{array}{cccc}
 5 & 1 & 1  & 1 \\
 1 & 5 & 1 & 1\\
 1 & 1 & 3 & -1\\
 1 & 1 & -1 & 3
\end{array}
\right)\, .
\end{align}
This state features neither quantum nor classical $\sigma_z$ correlations.
Within the interval $\theta \in [0,\pi)$ there are nine values of $\theta$ which can give rise to resonances (omitting the trivial resonance $\theta = \pi$).
For $\theta^\star \in \{\pi/5, 2\pi/5, 3\pi/5, 4\pi/5 \}$ we have $e^{i \theta^\star \lambda_3} = e^{i\theta^\star \lambda_0}$.
The steady state in the weak-reset limit is then 
\begin{widetext}
\begin{align}
\lim_{r\to0}\varrho_{\rm ss}=
\frac{1}{16}\left(
\begin{array}{cccc}
 7-\sqrt{2} & 1 & \frac{1+i}{\sqrt{2}+2} & \frac{1-i}{\sqrt{2}+2} \\[.2cm]
 1 & \sqrt{2}+3 & \frac{1-i}{\sqrt{2}}+1+i & 1-i+\frac{1+i}{\sqrt{2}} \\[.2cm]
 \frac{1-i}{\sqrt{2}+2} & 1-i+\frac{1+i}{\sqrt{2}} & 3 & -1+i \sqrt{2} \\[.2cm]
 \frac{1+i}{\sqrt{2}+2} & \frac{1-i}{\sqrt{2}}+1+i & -1-i \sqrt{2} & 3
\end{array}
\right).
\end{align}
\end{widetext}
Such a state  does not show entanglement, as evidenced by      $\mathcal{C} = 0$ (cf. Fig.~\ref{fig:Tgate}(a)).
However, it features classical correlations, $C \approx 0.055$ (cf. Fig.~\ref{fig:Tgate}(c)), as well as genuine quantum correlations as witnessed by a finite value of the LQU, $\mathcal{U}\approx 0.11$ (cf. Fig.~\ref{fig:Tgate}(d)).

For $\theta^\star \in \{ \pi/4,\pi/2,3\pi/4\}$, the levels $\lambda_3$ and $\lambda_1$ are resonant, which gives rise to the steady state 
\begin{align}
\lim_{r\to0}\varrho_{\rm ss} =
 \frac{1}{8} \left(
\begin{array}{cccc}
 3 & 1 & 0 & 0 \\
 1 & 3 & 0 & 0 \\
 0 & 0 & 1 & -1 \\
 0 & 0 & -1 & 1
\end{array}
\right).
\end{align}
This state is neither entangled nor discorded and shows no classical correlations.
However, from Fig.~\ref{fig:Tgate} we see that, in particular for the concurrence, the values $\theta^\star = \pi/4,\ 3\pi/4$ show a qualitatively different behaviour.
Rather than a sudden death as a function of $r$, the entanglement of the mixed state continuously vanishes in the limit $r\to0$, see also Fig.~\ref{fig:Tgate}(b).
The value $\theta^\star = \pi/2$ is special since it also puts on resonance the levels $\lambda_2$ and $\lambda_0$ which results in $U(\pi/2) = \mathds{1} \otimes \sigma_x$. In this case, the dynamics consists of classical spin flips of the second qubit, which explains the absence of correlations. However, from Fig.~\ref{fig:Tgate}(d) we still see that an island of non-vanishing LQU is connected to this special point.

Finally, for $\theta^\star \in \{ \pi/3,2\pi/3\}$ the levels $\lambda_3$ and $\lambda_2$ are resonant.
The corresponding steady state  reads
\begin{widetext}
\begin{align}
 \lim_{r\to0}\varrho_{\rm ss} =
 \frac{1}{16}\left(
\begin{array}{cccc}
 7-\sqrt{2} & 1 & \frac{1-i}{\sqrt{2}+2} & \frac{1+i}{\sqrt{2}+2} \\[.2cm]
 1 & \sqrt{2}+3 & 1-i+\frac{1+i}{\sqrt{2}} & \frac{1-i}{\sqrt{2}}+1+i \\[.2cm]
 \frac{1+i}{\sqrt{2}+2} & \frac{1-i}{\sqrt{2}}+1+i & 3 & -1-i \sqrt{2} \\[.2cm]
 \frac{1-i}{\sqrt{2}+2} & 1-i+\frac{1+i}{\sqrt{2}} & -1+i \sqrt{2} & 3
\end{array}
\right).
\end{align}
\end{widetext}
As in the first case, we observe absence of  entanglement, i.e., $\mathcal{C} =0$, a non-vanishing LQU of $\mathcal{U} \approx 0.11$ and finite classical correlations $C\approx 0.055$.

The structure of correlations beyond the weak-reset limit is shown in Fig.~\ref{fig:Tgate}.
Apart from the  fingerprints of the resonances in all investigated quantities, we see that portions of non-vanishing entanglement, LQU and correlations are connected, e.g., to the resonance at $4\pi/5$. 
The area of non-vanishing concurrence eventually broadens upon increasing $r$, giving rise to an entangled steady-state regime in an otherwise separable part of parameter space.
Remarkably, it appears to be quite  intricate to generate steady-state entanglement, as evidenced by large portions in parameter space of vanishing concurrence.
Nonetheless, correlations of quantum nature, as detected by the LQU, are much more extended.

\section{Non-Poissonian Resetting}
\label{sec:nm}

We now consider the case of non-Poissonian stochastic resetting (i.e., of a reset probability $r_n$ which is not constant) and address the question whether or not a steady state exists for the discrete-time process.
There are obvious instances in which the system will never relax to a steady state, e.g., $r_n = \delta_{n,\ell}$,  which was studied  in~\cite{Yin23}. On the other hand, we observe that for cases in which the values $r_n$ remain non-zero, for all $n$, a steady state is typically approached.  
In the following, by considering a specific example, we want to shed light on properties of  $r_n$ which may hinder relaxation to a steady state. 
Physically, it is intuitive that general conditions on the approach to a steady state should be found in the properties of the reset probabilities since the underlying dynamics is unitary and does not lead to a steady state, at least for finite systems.

To provide a quantitative measure of the approach of the system to a steady state, we  define the finite difference $\Delta \varrho (t) = \varrho(t+1) - \varrho(t)$.
Decomposing the system states into the eigenbasis of the unitary gate, we can write
\begin{widetext}
\begin{align}
\label{eq:drho}
 \Delta\varrho(t)
 &= \sum_{ij}\left[
 P_{t+1}(t+1) \left(\frac{u_j}{u_i}\right)^{t+1} +
 \sum_{n=0}^t \Delta P_n(t)\left(\frac{u_j}{u_i}\right)^{n}
 \right]\braket{e_i|0}\!\braket{0|e_j} \ket{e_i}\!\bra{e_j}
\end{align}
\end{widetext}
with $\Delta P_n(t) = P_n(t+1) - P_n(t)$.
Clearly, $\operatorname{tr} \Delta \varrho (t) = 0$ which yields the probability conservation condition
\begin{align}
\label{eq:proba_con}
 0 &= P_{t+1}(t+1)  + \sum_{n=0}^t \Delta P_n(t) .
\end{align}
This condition, combined with the necessity that $\lim_{t\to\infty}\Delta \varrho(t) = 0$, if a steady state exists, allows us to formulate the necessary and sufficient condition for $\varrho$ to reach a stationary state.
If $\braket{u_i|0}\!\braket{0|u_j} \neq 0$, then we must have 
\begin{align}
\label{eq:drho1}
\lim_{t\to\infty}\sum_{n=0}^t \Delta P_n(t)  \left[
 \left(\frac{u_j}{u_i}\right)^{n} - \left(\frac{u_j}{u_i}\right)^{t+1}
 \right]= 0
\end{align}
for all $i,j$.
This relation is rather involved  and can, in principle, be satisfied in a variety of ways.
To gain some understanding, let us  assume for the moment that the dynamics reaches a steady state.
This means the probability to observe a reset event becomes time-independent at long times, i.e.,  $\lim_{t\to \infty } P_0(t) = P_0^*$ and $\lim_{t\to \infty }\Delta P_0(t) = 0$.
In this case, (\ref{eq:recPn})  immediately implies that all $\lim_{t\to\infty}P_n(t) = P_n^*$ and (\ref{eq:proba_con}) then yields $\lim_{t\to \infty } P_t(t) = 0$. 
These relations seem to suggest that, in order not to approach a steady state, the stochastic resetting  should become irrelevant in the long-time limit. This can happen,  for instance, when $r_n$ decays sufficiently fast with $n$ (time from the previous reset). 
In this case, one could observe a large portion of trajectories that, if not resetting in the first few steps will never reset and would thus simply evolve unitarily.

To explore this intuition further, we now consider  a specific example and take a reset probability decaying as 
\begin{align}
 r_n = \frac{\gamma}{(n+1)^\alpha} ,
 \label{eq:decay-r}
\end{align}
where  $\alpha>0$ is the power-law exponent and $\gamma\in[0,1]$. 
From the reset probability, we can evaluate the probability for no reset  event to occur, $P_t(t)$, as
\begin{align}
 P_t(t) = \prod_{j=0}^{t-1} (1-r_j).
 \label{eq:Ptt}
\end{align}
This quantity behaves, for large $t$, as
\begin{align}
  P_t(t) \simeq
 \begin{dcases}
  \exp\left(-\gamma\zeta(\alpha)\right), \ &\text{if} \ \alpha >1\\[.25cm]
  t^{-\gamma}, \ &\text{if} \ \alpha =1\\
  \exp \left(-\frac{\gamma t^{1-\alpha}}{1-\alpha} \right), \ &\text{if} \ \alpha <1
 \end{dcases}\ .
 \label{eq:aympPtt}
\end{align}
A detailed derivation of this result can be found in  Appendix~\ref{app:Ptt}.
Importantly, we see that for $\alpha>1$ no steady state should be expected, since $P_t(t) \to {\rm const.}>0$. 
This implies that there is a finite probability to see a trajectory which never resets and thus undergoes unitary dynamics. 
For $0<\alpha<1$, we instead observe that $P_t(t) \to 0$ exponentially. This indicates that the probability of not observing reset events decays very rapidly. In these cases, one would expect the system to converge to a steady state. 
Interestingly, for  $\alpha = 1$ we still see that $P_t(t)$ tends to zero, however the decay is polynomial with $t$. 
Also in this situation, one may expect that a steady state is approached in the long-time limit. 

\begin{figure}[t!]
  \centering
  \includegraphics[width=0.49\columnwidth]{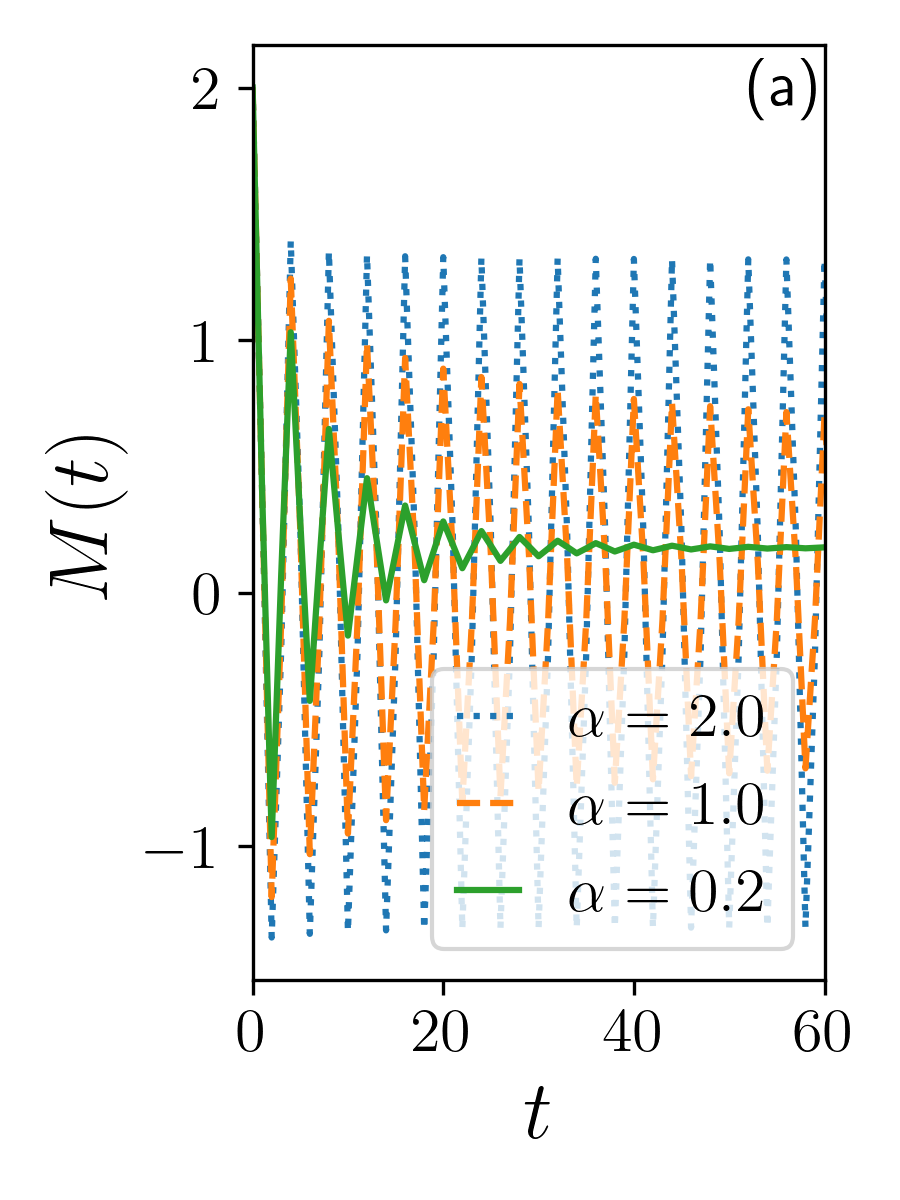}
  \includegraphics[width=0.49\columnwidth]{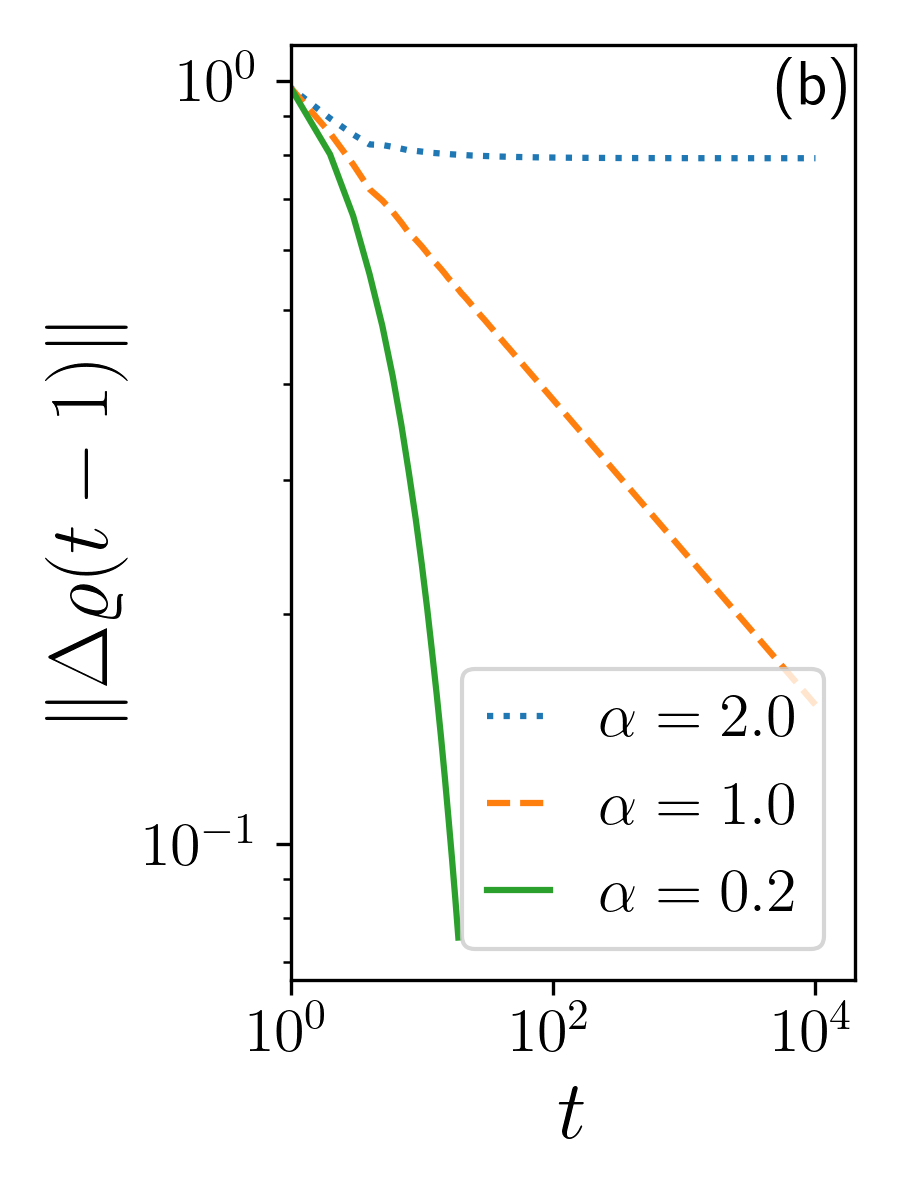}
  \caption{{\bf Relaxation to the steady state. } (a) Magnetization $M(t) = \operatorname{tr}[\varrho(t) (\sigma_z^{(1)} + \sigma_z^{(2)})]$ of a two-qubit system, subject to a stochastic resetting with probability as in (\ref{eq:decay-r}),  with $\alpha = 0.2, 1, 2$ and $\gamma = 0.2$.
  The unitary evolution is generated by the noninteracting discrete-time dynamics in (\ref{eq:NI}) with $\theta=\pi/4$. (b) Behavior of the norm of $\Delta\varrho(t-1)$ as a function of $t$ for the same process as in (a). The plot shows that, for $\alpha\le 1 $, $\|\Delta\varrho(t-1)\|\to 0$ as $t\to\infty$, indicating the existence of a steady state in this regime.  } 
  \label{fig:NP}
\end{figure}

To test these conclusions, we investigate the behavior of the finite difference $\Delta\varrho(t)$ for a two-qubit system undergoing the dynamics in \eqref{eq:NI} and subject to stochastic resetting with the reset probability in \eqref{eq:decay-r}. 
The results are shown in Fig.~\ref{fig:NP}, which shows indeed that a steady state is approached for $\alpha\le 1$.

\section{Conclusion}
\label{sec:conclusion}

We have considered a generic discrete-time quantum dynamics subject to a stochastic resetting process. 
For Poissonian reset events, we have derived the steady state of the system and analyzed its structure in the weak-reset limit. 
In general, the steady-state density matrix is (in the absence of degenerate eigenvalues of the generator of the unitary discrete-time dynamics) diagonal in the generator's eigenbasis.
However, we have shown that, for particular values of the gate parameter $\theta$, certain eigenvalues of the unitary operator can become ``resonant". 
Contrary to what happens for continuous-time reset processes, this results in off-diagonal contributions to the steady-state density matrix even in the absence of degenerate eigenvalues in the generator of the dynamics. 

By considering two specific examples, we have explored  correlations in the steady state of the discrete-time reset   process. 
We have demonstrated the existence of quantum and classical correlations, which can  be generated by the presence of the reset even in an otherwise noninteracting dynamics. 
Moreover, we have shown that the derived resonances can dramatically affect the structure of the steady state. 
Finally, we have focused on a non-Poissonian stochastic resetting and discussed conditions on the existence of the steady state for the process.
We have shown that if the reset probability decays sufficiently fast as a function of the time elapsed from the last reset event, then the process can feature trajectories which never reset and thus no steady state is reached.

Our work paves the way to further explorations of discrete-time resetting processes.
It would be particularly interesting to better understand the effects of non-Poissonian resets of quantum dynamics.
Another natural extension is to study our protocol in many-body systems and/or considering underlying dissipative dynamics.
Even a better understanding of the correlations in the steady state of  a three-qubit system seems already an interesting area for exploration.

Our findings shed light on  dynamical and  steady-state properties of discrete-time evolutions subject to random resets. In particular, they  highlight the emergence of resonances, not possible in continuous-time reset processes, which are responsible for rather abrupt changes in the properties of the quantum state describing the process. 
Given that discrete-time reset dynamics can nowadays be realized on quantum computers, our results may be useful to engineer steady states with  controllable properties.

\section*{Acknowledgement}
The authors would like to thank Adam Lowe for useful discussions that helped advance this work. FC acknowledges funding from the Deutsche Forschungsgemeinschaft (DFG, German Research Foundation) through the Research Unit FOR 5413/1, Grant No.\ 465199066 as well as from the European Union’s Horizon Europe research and innovation program under Grant Agreement No.\ 101046968 (BRISQ). FC is indebted to the Baden-Württemberg Stiftung for the financial support by the Eliteprogramme for Postdocs.

\appendix

\section{Numerical details}
\label{app:numerics}
In this appendix we briefly sketch how we have obtained the numerical results in the paper.
The results in the Poissonian and non-Poissonian scenario are generated from different codes and thus, we shall discuss them in separate subsections.

\subsection{The Poissonian case}

For the numerical study, it is convenient to reformulate the discrete time resetting dynamics in vectorized form by introducing the doubled Hilbert space $\mathcal{H}_{2} = \mathcal{H}\otimes\mathcal{H}$.
Linear forms $\varrho$ on $\mathcal{H}$ are then mapped to vectors $\kket{\varrho}$ in $\mathcal{H}_2$.
The mapping is given by
\begin{align}
    \kket{\varrho(t)} = \sum_{ij}\bra{i}\varrho(t)\ket{j} \ket{i}\ket{j}.
\end{align}
For a Kraus map $\varrho(t+1) = \sum_\mu K_\mu \varrho(t) K_\mu^\dagger$ we can then derive a time evolution of the vectorized form, viz.,
\begin{align}
    \kket{\rho(t+1)}
    &= \sum_\mu K_\mu \otimes K_\mu^* \kket{ \rho(t)} \equiv
    \mathcal{K} \kket{\rho(t)}.
    \label{eq:ds-rec}
\end{align}
In the Poissonian cases we consider, the Kraus map is given by
\begin{align}
 \mathcal{K}   =(1-r) U \otimes U^* + r\kket{\varrho(0)}\bbra{\mathds{1}}.
\end{align}
The discrete time Poissonian resetting dynamics is then given by
\begin{align}
\kket{\varrho(t)} = \mathcal{K}^t \kket{\varrho(0)}.
\end{align}
It is easy to derive the closed form for $\mathcal{K}^t$ by mathematical induction.
It is given by
\begin{align}
\begin{split}
    \mathcal{K}^t =& (1-r)^t (U\otimes U^*)^t \\
    &+ r \frac{1-(1-r)^t (U\otimes U^*)^t}{1-(1-r) (U\otimes U^*)} \kket{\varrho(0)}\bbra{\mathds{1}}.
\end{split}
\end{align}
This readily allows the evaluation of the stationary state in the limit $t\to\infty$, viz.,
\begin{align}
\mathcal{K}^t \to r \frac{\kket{\varrho(0)}\bbra{\mathds{1}}}{1-(1-r) (U\otimes U^*)}, \ \text{for}\ t\to\infty.
\end{align}
This allows us to write the steady state $\kket{\varrho_{\rm ss}}$ as
\begin{align}
    \kket{\varrho_{\rm ss}} =  \frac{r}{1-(1-r) (U\otimes U^*)} \kket{\varrho(0)}.
\end{align}
Hence, to determine the exact steady state it suffices to invert the operator $1-(1-r) (U\otimes U^*)$.

\subsection{The non-Poissonian case}
In the non-Poissonian case, it is no longer possible to formulate the dynamics of the mixed state through a simple recursion relation as (\ref{eq:ds-rec}).
This is because different components of the mixture are subject to different reset probabilities as can be seen in Fig.~\ref{fig:tree}.
Furthermore, as we discussed in Sec.~\ref{sec:nm}, there may not even be a well-defined steady state.
Therefore, we explicitly evolve the mixed state for each time step by establishing an array in which all states $\ket{n_\theta}$ are saved that contribute to the mixture up to the current simulation time.
In a second array, we track the corresponding probabilities $P_n(t)$ which are propagated in time following (\ref{eq:recPn}).
At each time step $t$ we then construct the corresponding mixed state $\kket{\varrho(t)}$ and compare it with the state $\kket{\varrho(t-1)}$ at the previous time step using the $1$-norm, i.e., $\epsilon = \left|\left|\kket{\varrho(t)} - \kket{\varrho(t-1)}\right|\right|_1$.
We then define a convergence threshold of $\epsilon = 10^{-10}$ which we try to achieve within $10,000$ time steps.

\section{Correlation function in the noninteracting case}
\label{app:correlator}

In this section we illustrate how to obtain the analytical expression for the spin correlation function of a non-interacting two qubit system.
Similar calculations can be carried out in the interacting case and are thus not further shown in this appendix.

The connected spin correlator reads
\begin{align}
C = \langle\sigma_z^{(1)} \sigma_z^{(2)}\rangle - \langle\sigma_z^{(1)}\rangle\langle\sigma_z^{(2)}\rangle.
\end{align}
For this section and the following ones it becomes convenient to define the function
\begin{align}
\begin{split}
f(r,x)&=r\sum_{j=0}^{\infty}(1-r)^j \cos\left( jx \right)\\
&=r\frac{1-(1-r)\cos(x)}{1+(1-r)^2-2(1-r)\cos(x)}.
\end{split}
\end{align}
We will be using $\langle\sigma_z^{(1)}\rangle= \langle\sigma_z^{(2)}\rangle= \langle\sigma_z\rangle$ since both spins are in the same initial state and subject to the same dynamics, and are thus indistinguishable.
We calculate $\left<\sigma_z\right>$ via
\begin{align}
\langle\sigma^z\rangle=r\sum_{j=0}^{\infty}(1-r)^j \cos\left( 2j\omega\right)
= f(r,2\omega).
\end{align}
Similarly, we may evaluate the two point spin correlator, viz.,
\begin{align}
&\langle\sigma_z^{(1)} \sigma_z^{(2)}\rangle
=
\frac{1+f(r,4\omega)}{2}.
\end{align}
Hence, the spin correlation function reads
\begin{align}
C(r,\omega)=
\frac{1+f(r,4\omega)}{2}
-
\left[f(r,2\omega)\right]^2.
\end{align}
\vspace{.2cm}

\section{Asymptotic analysis of no-reset probability}
\label{app:Ptt}
Here, we illustrate the calculation that yields the asymptotic behavior of $P_t(t)$, indicated in (\ref{eq:aympPtt}) for the reset rates from (\ref{eq:decay-r}).
We recall (\ref{eq:Ptt}) where $P_t(t)$ is indicated as a product, i.e.,
\begin{align}
 P_t(t) = \prod_{j=0}^{t-1} (1-r_j).
 \label{eq:Ptt-app}
\end{align}
First, we consider the logarithm of (\ref{eq:Ptt-app}) since this transforms the sum into a product.
We obtain
\begin{align}
 \log P_t(t) = \sum_{j=0}^{t-1} \log(1-r_j) .
\end{align}
Next, we approximate $\log(1-r_j)\simeq -r_j$.
This is true as long as the $r_j$ decay with increasing index, i.e., $r_j \to 0$ for $j\to \infty$ and if we are only interested in the asymptotic behavior of $P_t(t)$.
This approximation then yields
\begin{align}
 \log P_t(t) \simeq - \sum_{j=0}^{t-1} \frac{\gamma}{(j+1)^\alpha}.
\end{align}
The further analysis is distinct for $\alpha = 1$ and therefore we treat it separately.

{\it The case $\alpha \neq 1$}.
In this case, we can directly evaluate the sum and the result is readily expressed in terms of the Hurwitz Zeta function $\zeta(s,a)$ and the Zeta function $\zeta(s)$.
We find
\begin{align}
 \log P_t(t) \simeq - \gamma \left( \zeta(\alpha) - \zeta(\alpha,1+t) \right).
 \label{eq:logPtt}
\end{align}
For large $t$ we may replace the Hurwitz Zeta function by its asymptotic form and reveal
\begin{align}
 \log P_t(t) \simeq -\gamma\left(\zeta(\alpha) + \frac{t^{1-\alpha}}{1-\alpha}\right).
\end{align}
Thus we see that the behaviors for $\alpha >1$ and $\alpha<1$ strongly differ.
We find the results stated in the main text:
\begin{align}
 P_t(t) \simeq
 \begin{dcases}
  \exp(-\gamma \zeta(\alpha)) &\text{if} \quad \alpha> 1;\\[.5cm]
  \exp\left(- \frac{\gamma t^{1-\alpha}}{1-\alpha}\right) &\text{if} \quad \alpha< 1.
 \end{dcases}
\end{align}

{\it The case $\alpha = 1$}.
(\ref{eq:logPtt}) clearly highlights that there is a need for a different resummation for $\alpha =1$.
We find
\begin{align}
 \log P_t(t) \simeq -\gamma \left(\gamma_E  +  \psi (t+1)\right),
\end{align}
where $\gamma_E$ is Euler's constant $\gamma_E \approx 0.577216$ and $\psi(z)$ is the Polygamma function.
The leading asymptotic behavior of the Polygamma function is
\begin{align}
 \psi(t) \simeq \log(t).
\end{align}
This implies that the asymptotic behavior of $P_t(t)$ is algebraic, i.e.,
\begin{align}
 P_t(t) \simeq e^{-\gamma \gamma_E} t^{-\gamma} .
\end{align}

\bibliography{refs}
\bibliographystyle{quantum}
\end{document}